\documentclass{article}

\usepackage{PRIMEarxiv}

\usepackage[utf8]{inputenc} 
\usepackage[T1]{fontenc}    
\usepackage{hyperref}       
\usepackage{url}            
\usepackage{booktabs}       
\usepackage{amsfonts}       
\usepackage{nicefrac}       
\usepackage{microtype}      
\usepackage{lipsum}
\usepackage{fancyhdr}       
\usepackage{lmodern} 
\usepackage{graphicx}
\usepackage[numbers,square,sort&compress]{natbib}
\usepackage[font=small]{caption}
\usepackage[position=top,labelformat=empty]{subcaption}
\usepackage[export]{adjustbox}
\usepackage{chemformula}

\title{Advancing Wavefront Shaping with Resonant Nonlocal Metasurfaces: Beyond the Limitations of Lookup Tables
}

\author{
  Enzo Isnard \\
  THALES Research \& Technology, 1 avenue Augustin Fresnel, 91120 Palaiseau, France\\
  Universit\'{e} C\^{o}te d'Azur, Inria, CNRS, LJAD, 06902 Sophia Antipolis Cedex, France \\
  \And
  Sébastien Héron \\
  THALES Research \& Technology, 1 avenue Augustin Fresnel, 91120 Palaiseau, France\\
   \And
  Stéphane Lanteri , Mahmoud Elsawy* \\
  Universit\'{e} C\^{o}te d'Azur, Inria, CNRS, LJAD, 06902 Sophia Antipolis Cedex, France \\
  \texttt{mahmoud.elsawy@inria.fr} \\
}

\begin{document}
\maketitle

\begin{abstract}
	Resonant metasurfaces are of paramount importance in addressing the growing demand for reduced thickness and complexity, while ensuring high optical efficiency. This becomes particularly crucial in overcoming fabrication challenges associated with high aspect ratio structures, thereby enabling seamless integration of metasurfaces with electronic components at an advanced level. However, traditional design approaches relying on lookup tables and local field approximations often fail to achieve optimal performance, especially for nonlocal resonant metasurfaces. In this study, we investigate the use of statistical learning optimization techniques for nonlocal resonant metasurfaces, with a specific emphasis on the role of near-field coupling in wavefront shaping beyond single unit cell simulations. Our study achieves significant advancements in the design of resonant metasurfaces. For transmission-based metasurfaces, a beam steering design outperforms the classical design by achieving an impressive efficiency of 80\% compared to the previous 23\%. Additionally, our optimized extended depth-of-focus (EDOF) metalens yields a remarkable five-fold increase in focal depth, a four-fold enhancement in focusing power compared to conventional designs and an optical resolution superior to 600 cycle/mm across the focus region. Moreover, our study demonstrates remarkable performance with a wavelength-selected beam steering metagrating in reflection, achieving exceptional efficiency surpassing 85\%. This far outperforms classical gradient phase distribution approaches, emphasizing the immense potential for groundbreaking applications in the field of resonant metasurfaces.
\end{abstract}


	\section{Introduction} 
	
	Metasurfaces are flat optical components that have the ability to alter electromagnetic wave properties, including phase, amplitude, and polarization \cite{sc2011,genevet2017recent,lalanne_98blazed,Lin_Hasman_Science2014}. They are constructed using subwavelength-sized elements known as meta-atoms that interact with incoming waves to achieve a specific outcome. 
	In this context, several applications have already been proposed such as ultra-compact metasurface-based deflectors, beam splitters, lenses, waveplates, holograms \cite{elsawy2021multiobjective,metalens2018broadband,metalenschen2018broadband,metalenspan2022dielectric,hologramsong2020ptychography,emittxie2020metasurface}. 
	
	In metasurface design, local and nonlocal effects refer to the way the properties of individual resonators affect the behavior of the whole structure and especially of their neighbors. A local metasurface is a type of metasurface in which the individual meta-atom only interacts with external electromagnetic waves that are incident directly on them. In other words, each resonator acts independently, and the electromagnetic response of the metasurface can be determined by a simple lookup table that relates the properties of the resonators to the desired wavefront. Generally speaking this approach could be applied for metasurfaces based on the modification of the effective refractive index of low-quality factor truncated nano-waveguides \cite{lalanne1999designgratings,propagation2015broadband, lalanne_98blazed,metalens2018broadband}. 
	It is also applicable to metasurfaces which rely on the Pancharatnam Berry (PB) phase induced by birefringent meta-atoms. 
	\cite{HasmanOL2002PBB,PBB2014silicon, PBB2016metalenses,PBBtian2020manipulation}.
	
	On the other hand, nonlocal metasurface is a type of metasurface in which the meta-atoms cannot be considered isolated and the response of each meta-atom depends on the state of its neighbors \cite{Alu_Nonlocalmultifunctional,caiNonlocalinverse}. In other words, in a nonlocal metasurface, the meta-atoms are not only interacting with the external waves that are incident directly on them, but also with the waves that are incident on their neighboring meta-atoms. This metasurface interacts with light in a way that is not local, meaning that the effect of meta-atoms on the light is not limited to their immediate vicinity. Instead, the behavior of the entire metasurface is determined by the collective behavior of all the meta-atoms, making it difficult to predict the response using a lookup table from the unit cell simulations.
	
	The nonlocal interaction is inherently present for metasurface considering phase addressing mechanism relies on resonant scattering of light, and the associated light scattering modulation properties \cite{kivsharliu2018generalized,Hygens_yu2015high,plasmonic_huang2013three,plasmonic_bin2021ultra}. Recent works based on lookup table considerations revealed an intriguing connection between the full phase modulation and the topological effects governing the resonant light interaction of nanoparticles \cite{colom2023crossing,elsawyGT2022}.
	
	Furthermore, the concept of nonlocal metasurfaces has been extended to structures whose governing principles are deeply rooted in the symmetry features of quasi-bound states in the continuum (q-BIC) \cite{Alu_Nonlocalmultifunctional,Alu2020selection,Alu2021active} for selective wavefront control. Yet, the overall efficiency (ratio of the power in the desired direction compared to the incident power) is modest since most of theses studies predict the overall performance response using a simple lookup table from the single-unit cell simulation which neglect the cross talk between adjacent resonant elements.
	
	The importance of resonant nonlocal metasurfaces becomes evident in their ability to facilitate a remarkable reduction in metasurface thickness and meta-atom complexity. This reduction in structural requirements is crucial for several reasons. First, it allows for the integration of metasurfaces within existing electronic components, paving the way for advanced functionalities and enhanced performance in integrated photonic systems. Second, the decreased thickness simplifies the fabrication process, enabling large-scale manufacturing with higher precision and reduced costs. 
	
	In this study, we demonstrate that the local approach is not suitable  to predict the performance of highly resonant metasurfaces. Moreover, we propose an optimization method for nonlocal metasurfaces based on a global statistical learning optimization method referred as EGO (Efficient Global Optimization) which has been successfully applied to local metasurfaces \cite{elsawy2019global,elsawy2021multiobjective,elsawy2021optimization}.
	
	EGO is based on building a surrogate model of the objective function using a Gaussian process regression (GPR) approach. This model is used to approximate the true objective function and to guide the search for the optimal solution. EGO iteratively updates the surrogate model by adding new observations to it, and then uses an acquisition function to select the next point to evaluate. The acquisition function balances exploration and exploitation to identify the next point that is most likely to improve the objective function's value. EGO has several advantages over other optimization algorithms. First, the surrogate model that is built by GPR is capable of capturing complex and nonlinear relationships between input and output variables. 
	Second, EGO balances exploration and exploitation, enabling it to efficiently search the design space for the global optimum while avoiding getting trapped in local optima.
	
	One of the key strengths of EGO is its ability to handle expensive-to-evaluate black-box functions. In such cases, evaluating the objective function is computationally intensive, and traditional optimization methods may be too slow to converge to a satisfactory solution. EGO mitigates this by using a surrogate model to approximate the objective function, reducing the number of function evaluations required to find the global optimum. By using a Gaussian process regression surrogate model and a carefully designed acquisition function, EGO efficiently explores the design space to identify the global optimum while minimizing the number of function evaluations required. Its ability to balance exploration and exploitation and handle expensive black-box functions make it a popular choice for a wide range of optimization problems.
	
	We demonstrate the effectiveness of our method by applying it to the design of a nonlocal metasurface for wavefront control with various mechanisms ranging from classical Huygens's metasurface \cite{kivsharliu2018generalized} to wavelength selective wavefront metagratings. The results show that our method is able to achieve superior performance compared to traditional synthesizing approach based on locality assumption, with a significant reduction in the number of function evaluations compared to conventional global optimization methods \cite{caiNonlocalinverse}. Overall, our study provides a promising approach for the optimization of nonlocal metasurfaces, and highlights the potential of EGO method for solving complex electromagnetic design problems.
	\section{Huygens's metasurface}
	\label{subsec:Hugens_discc}
	
	Huygen's metasurface is a promising approach to reduce the aspect ratio of constitutive elements of metasurfaces, which is crucial to reduce their fabrication complexity.
	It is a type of resonant metasurface that is designed to manipulate the scattering of light in a highly directional manner. It consists of an array of subwavelength resonators that are carefully engineered to exhibit both magnetic and electric dipole resonances at the same frequency. By controlling the relative phase and amplitude of the magnetic and electric dipoles in each resonator, the metasurface can achieve the required full phase modulation associated with near-unity transmission amplitude \cite{kivshar2020dielectric,kivsharliu2018generalized,Kivshar2017}.

	The Huygens's metasurface configuration has diverse potential applications, including optical communication, imaging, and sensing. However, the design of such metasurfaces is challenging due to the intricate interplay between the magnetic and electric dipole resonances, which is determined by the position of the poles and zeros as elaborated in Ref.\cite{colom2023crossing}. Moreover, the nanoresonators have extended evanescent electromagnetic fields that couple their resonances beyond adjacent neighbors, making the conventional design of phase-gradient metasurfaces complicated. In addition, various factors, such as the geometrical parameters of the resonators, the polarization and wavelength of the incident light, and the surrounding environment, can affect the complex interplay between the resonances. Therefore, designing a high-performance Huygen's metasurface necessitates meticulous consideration of these factors and advanced optimization techniques to achieve the desired properties.
	
	\subsection{Huygen's metasurface for beam steering}
	\label{subsec:steering}
	
\begin{figure*}[htb!]
		\centering
		\includegraphics[width=1.0\textwidth]{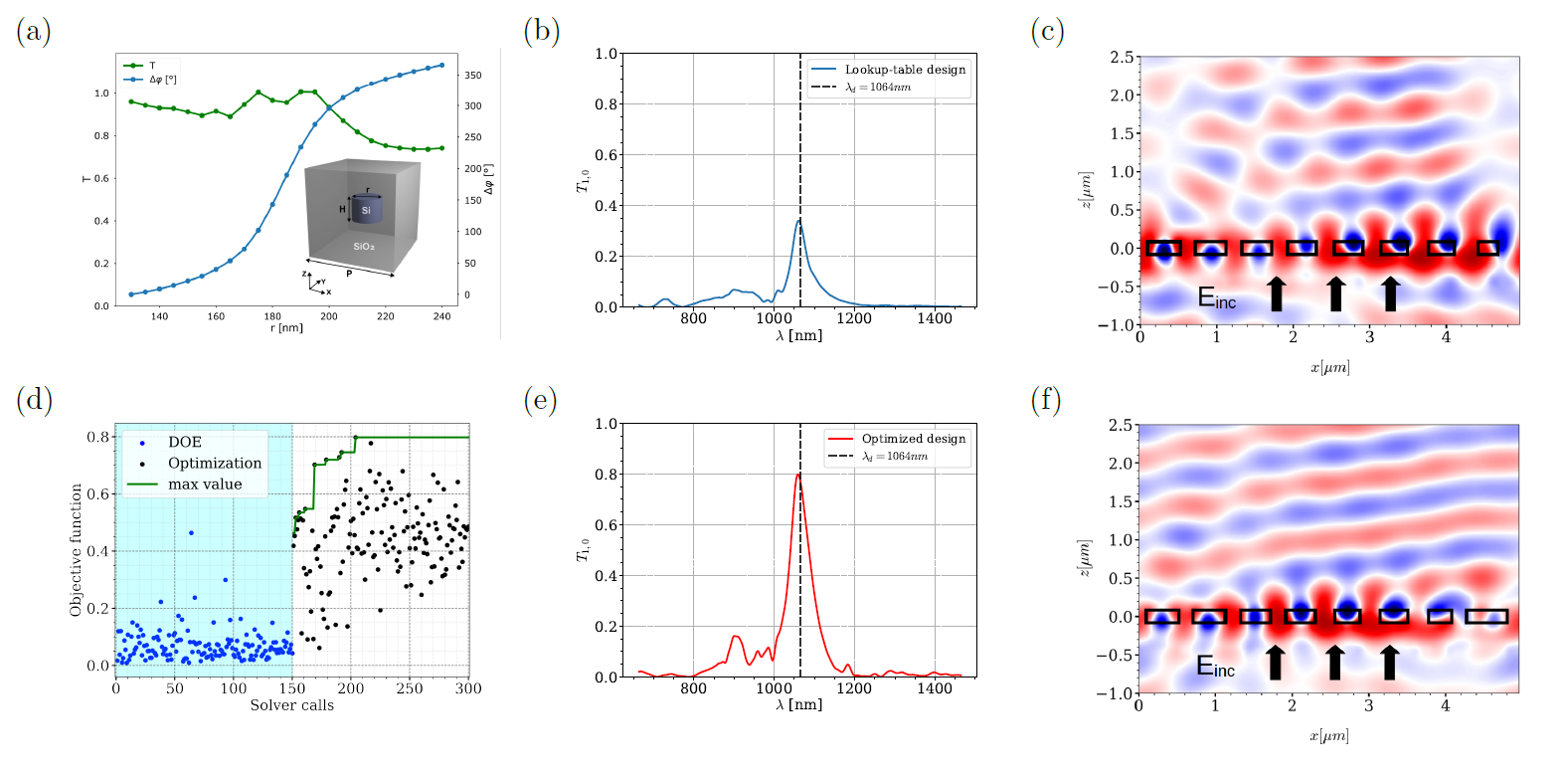}
		
		\caption{Optimization results for resonant Huygens's metasurface. (a): refers to phase and transmission response for single unit-cell simulations as a function of the radius of the silicon pillar at $\lambda=1064$ nm. The inset in (a) indicates the schematic representation of the structure under consideration. We consider a silicon pillar encapsulated in \ch{SiO2} material. The height of the pillars and the period are fixed as $h=170$ nm and $P=620$ nm. The radius of the pillars can vary from 130nm to 240nm. (b-c) beam steering configuration based on classical synthesizing approach where the radii of the cylinders are varying in a gradient manner based on the look-up table given in (a), for a fixed period $P=620$ nm. The field map represents $\Re(E_y)$ at $\lambda=1064$ nm.
			(d) optimization results for 8 pillars by varying the radii of each pillar and the distances in between them (15 parameters). The blue points refer to the design of experiment phase (DOE), the black points indicate the EGO iterations. The green curve is for the best values during the optimization process. (e) Power deflected in the first order mode for the optimal design obtained from the EGO, the corresponding field profile of $\Re(E_y)$ is given in (f).}
		\label{fig1:Hygen's}
	\end{figure*}

	To demonstrate the effectiveness of our statistical learning optimization method in designing efficient nonlocal metasurfaces, we first consider the resonant structures presented in \cite{ozdemir2017polarization} to design beam steering device. The structure is composed of silicon nanocylinders with a refractive index of $n=3.56$ embedded in a host silica with a refractive index of $n=1.45$, as shown in the inset in Fig.\ref{fig1:Hygen's}(a), with a fixed height of $h=170$ nm.
	We initially follow the local approach design methodology by constructing a look-up table starting from a single unit-cell with periodic boundary conditions. The subwavelength period is fixed at $P=620$ nm. In Fig.\ref{fig1:Hygen's}(a), we vary the radius of the nano-resonator to achieve the desired response, i.e., full phase modulation (blue curve) associated with high-transmission response (green color) for a fixed wavelength of $\lambda=1064$ nm. To complete the first step, we designed a beam steering design by considering 8 unit cells (to sample 0-$2\pi$ phase) with a phase gradient distribution. We focus on the Transverse Electric (TE) polarized light case with the electric field  oscillating in the y-direction \cite{gigli2021fundamental} 
	based on this classical synthesizing phase approach, the maximum light deflection efficiency in the first-order mode does not exceed $35\%$ at $\lambda=1064$ nm (see blue curve in Fig.\ref{fig1:Hygen's}(a)). This is due to the strong near-field coupling related to the nonlocal coupling between the resonators, which is not taken into account in the classical design approach (see Fig.\ref{fig1:Hygen's}(c)).
	
	Unlike the classical design, in order to optimize the near-field coupling, the optimization scenario requires optimizing the radius of each resonator along with the distances between them. For the 8-pixels configuration, we optimize 15 parameters and the objective is to maximize light in the first order mode. This is a challenging optimization scenario, therefore, we rely on our advanced numerical methodology by considering high-level statistical learning optimization technique that requires less iterations compared to classical global methods \cite{elsawy2019global} together with a high-fidelity 3D electromagnetic solver from the DIOGENeS suite \cite{DIOGENES} to capture effectively the strong-near field coupling. The optimization results are depicted in Fig.\ref{fig1:Hygen's}(d). The blue points refer to the Design Of Experiment (DOE) iterations, while the black points indicate the the optimization phase. The green curve represents the best values found during the optimization process. For more details about the optimization methodology, we refer to our recent works \cite{elsawy2019global,elsawy2020Review,elsawy2021multiobjective,elsawy2021optimization}. The best design can reach a deflection efficiency of $80\%$ as it is indicated in Fig.\ref{fig1:Hygen's}(e). The huge improvement in the performance compared to the classical design is due to the near-field coupling, which is now taken into account in the optimization scenario as illustrated in the field profile depicted in Fig.\ref{fig1:Hygen's}(f).    
	\subsection{Extended depth of focus metalens (EDOF)}
	\label{subsec:EDOF}
	Metalenses have emerged as a highly promising technology in the field of optics, capable of effectively assuming the role of extended depth of focus (EDOF) lenses. 
	Diverging from traditional lenses, EDOF lenses exhibit a consistent point spread function (PSF) throughout an extended range along the optical axis. By leveraging EDOF lenses, not only can novel functionalities be realized, such as the ability to bring objects at different distances from the lens into sharp focus, but they also alleviate the stringent constraints associated with precisely aligning lenses on top of a sensor. Extending the DOF also increases the spectral operation range because it reduces the impact of the focal shift due to chromatic aberrations on the quality of the images obtained \cite{baranikov2023large}.
	
	Conventional techniques, including cubic, shifted axicon, log-asphere, and SQUBIC lenses, have been proposed for synthesizing phase masks to achieve EDOF properties \cite{EDOFhuang2020design}. However, these traditional synthesis approaches encounter difficulties in striking a balance between maximizing focal efficiency and extending the focal depth. The limitations of the lookup table approach become apparent in addressing the complex requirements of achieving the desired EDOF properties. Notably, inverse design methods based on traditional optimization mechanisms have been utilized to maximize EDOF metalens performance, particularly in relying on classical optimization algorithm \cite{EDOFzheng2022designing,EDOFxiao2023inverse} with numerical aperture not exceeding 0.44 accompanied with huge number of simulations.
	
	The proposed design for this application involves a cylindrical metalens composed of unit cells described in Fig.\ref{fig1:Hygen's}(a). The lens comprises 18 unit cells aligned along the x-axis with periodic boundary conditions along the y-axis to mimic 1D metalens. We start from a design with a NA of 0.6 and we extend the focal depth in a rectangle from the $f-\mathrm{DOF}$ to $f+5\mathrm{DOF}$ with a width of 2 $\mathrm{FWHM}$ where $f$ is the focal distance and DOF is given by the formula in \cite{EDOFbayati2020inverse}.
	
	In order to extend the depth of focus and achieve a uniform optical field distribution within the target depth, we define the objective function as the geometric mean of the mean power and the minimum power obtained from 20 equally spaced surfaces above and below the focal point. Using a minimum in the objective is primordial in order to obtain a design with a homogeneous intensity in the target region.
	This objective function allows us to simultaneously increase the focal depth along the z-axis, decrease the FWHM along the x-axis, and enhance the focusing power efficiency. 
	
\begin{figure*}[htb!]
		\centering
		\includegraphics[width=1.0\textwidth]{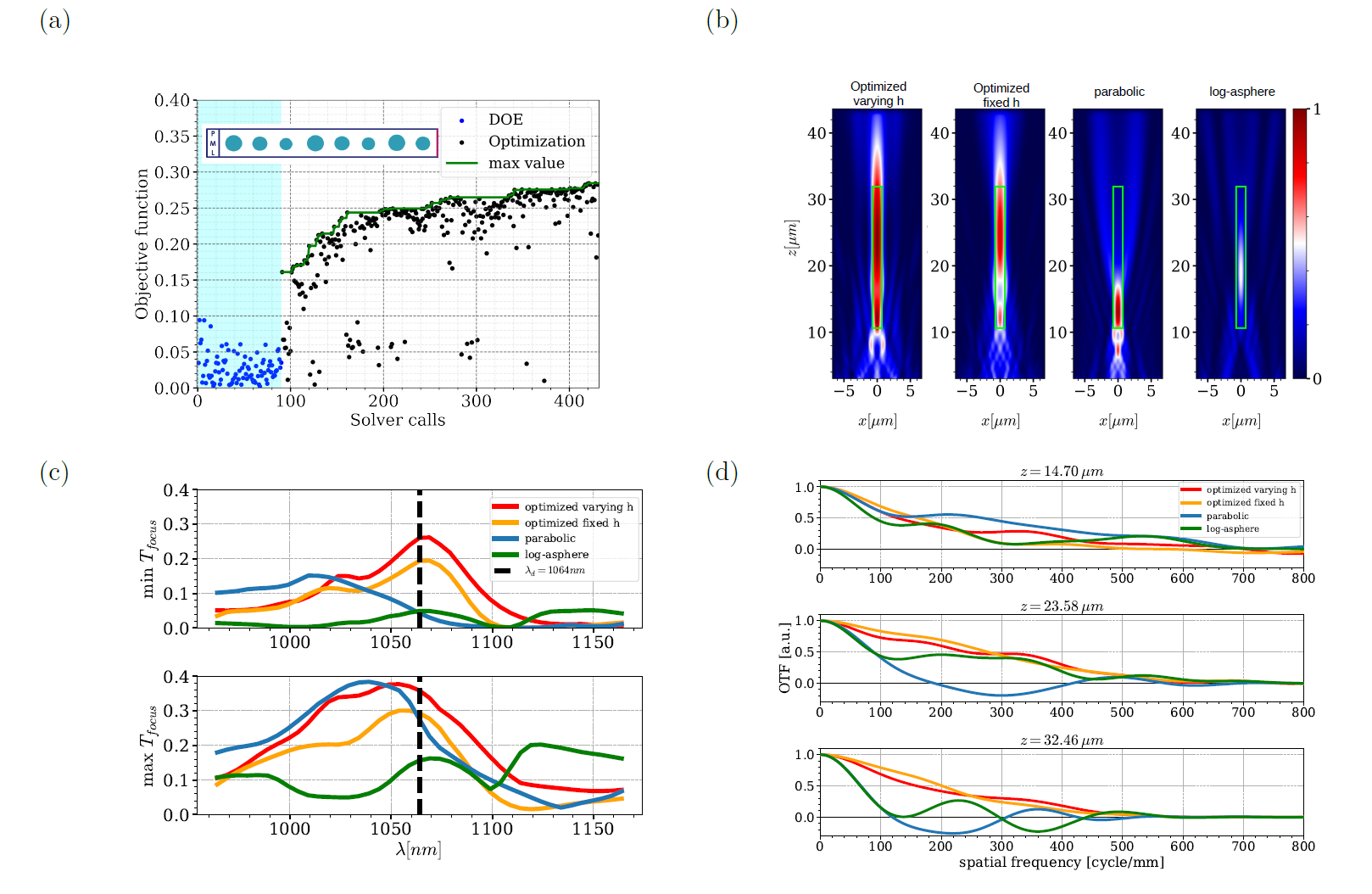}
		
		\caption[This is for my LOF]{Optimization results for the EDOF metalens. (a) Objective function values as a function of the number of simulations. The insight represents the geometry of half of the metalens in the $x$ direction. (b) Field maps of $\|\mathbf{E}\|^2$ at $1064$ nm of the optimized design with fixed and varying $h$, the parabolic designs and of the log-asphere design. The green rectangle represents the region in which we try to increase the DOF (c) Plot of the minimum and maximum focus efficiency over 20 equally spaced rectangles in the target region. The focus efficiency is defined as the part the incident power that is transmitted in a focus plane. The focus efficiency for the optimized design with a varying $h$ varies from 26\% to 35\% while for the parabolic designs it varies between 0.05\% and 27\%. (d)  Optical Transfer Functions (OTFs) of each designs at $f$, $f$ + 2.5 DOF and $f$ + 5 DOF. Optimized designs preserve spatial frequencies up to 600 cycle/mm at the three locations whereas the classical ones both fail at $f$ + 5 DOF.  }
		\label{fig2:metalens}
	\end{figure*}

	The employed geometric parametrization in this study aligns with the methodology presented in Section \ref{subsec:steering}, which encompasses the diameter and distances between the resonators. However, to enhance the degrees of freedom, we also explored the variation of the resonator height, ranging from 150 nm to 250 nm, resulting in the optimization of a total of 19 parameters. The optimization process, as depicted in Fig.\ref{fig2:metalens}(a), initiates with a Design of Experiments (DOE) consisting of 90 designs, indicated by the blue points. The process concludes after 450 simulations, represented by the black points in Fig.\ref{fig2:metalens}(a). The optimal design achieved a performance of 30\% according to our objective function. Furthermore, the electric field intensity $(\|\mathbf{E}\|^2)$ demonstrates an uniform distribution across the entire focal zone, as shown in the first sub-figure of Fig.\ref{fig2:metalens}(b).
	
	The optimization process converged to a specific height value $h=189.5$ nm, which differs from the height utilized in the lookup table presented in Fig.\ref{fig1:Hygen's}(a). To further investigate this aspect, we performed another optimization run while keeping the height fixed at $h=170$ nm, as defined in Fig.\ref{fig1:Hygen's}(a).
	
	The field profile distribution in the case with a constant height exhibits inferior performance compared to the case with varying height, as depicted in the first two sub-figures of Fig.\ref{fig2:metalens}(b). Moreover, the traditional parabolic phase configuration yields a narrow focal spot that fails to cover the full depth of focus. In order to provide a comprehensive comparison, the log-sphere synthesizing approach is also considered. The central anular zone and the outer-most one of the log-asphere phase profile are chosen to be respectively $f$ $-$ DOF and $f$ + 5 DOF. However, this approach only achieves an EDOF of $4$ $\mu m$ accompanied by a significant decrease in power efficiency, as indicated in Fig.\ref{fig2:metalens}(b). In conclusion, the optimized design with varying height demonstrates an EDOF of approximately $20$ $\mu m$, which is five times larger than that of the classical design. Additionally, Fig.\ref{fig2:metalens}(c) illustrates the enhanced focusing efficiency of the optimal design, surpassing that of the classical designs by a factor of three in average. This gain is greatly due to a large reduction of the back-scattering of the structure. 50\% of the losses of all designs are due to the small aperture of the lenses, which leads to unavoidable diffraction (see Supplementary Informations 1.4 for more details).
	
	Furthermore, in Fig.\ref{fig2:metalens}(d), we present the Optical Transfert Function (OTF) at three distinct z values: the focal point, the final positions within the focal depth zone and a position between the two. Notably, the optimized designs OTFs keep a similar shape at the three locations while the classical ones can not capture spacial frequencies higher than 100 cycle/mm at the second location. This implies that our design demonstrates significantly improved tolerance to changes in the focal plane position while maintaining nearly the same power efficiency and optical resolution. These results confirm that the local lookup table approach is unsuitable for complex metalens objectives due to the strong near-field coupling, which is a crucial factor for achieving the desired objectives. Importantly, our optimization process only required 450 iterations to converge to the globally optimal design, considering 19 parameters. This is made possible by the use of the EGO algorithm, which allows us to achieve convergence with a significantly lower number of iterations compared to standard state-of-the-art optimization algorithms \cite{EDOFzheng2022designing,EDOFxiao2023inverse}.
	\section{Nonlocal resonant metasurface in reflection}
	\label{sec:GT_gratings}
	\begin{figure*}[ht]
		\begin{center}
			\includegraphics[width=1.0\textwidth, trim={300 0 300 0}]{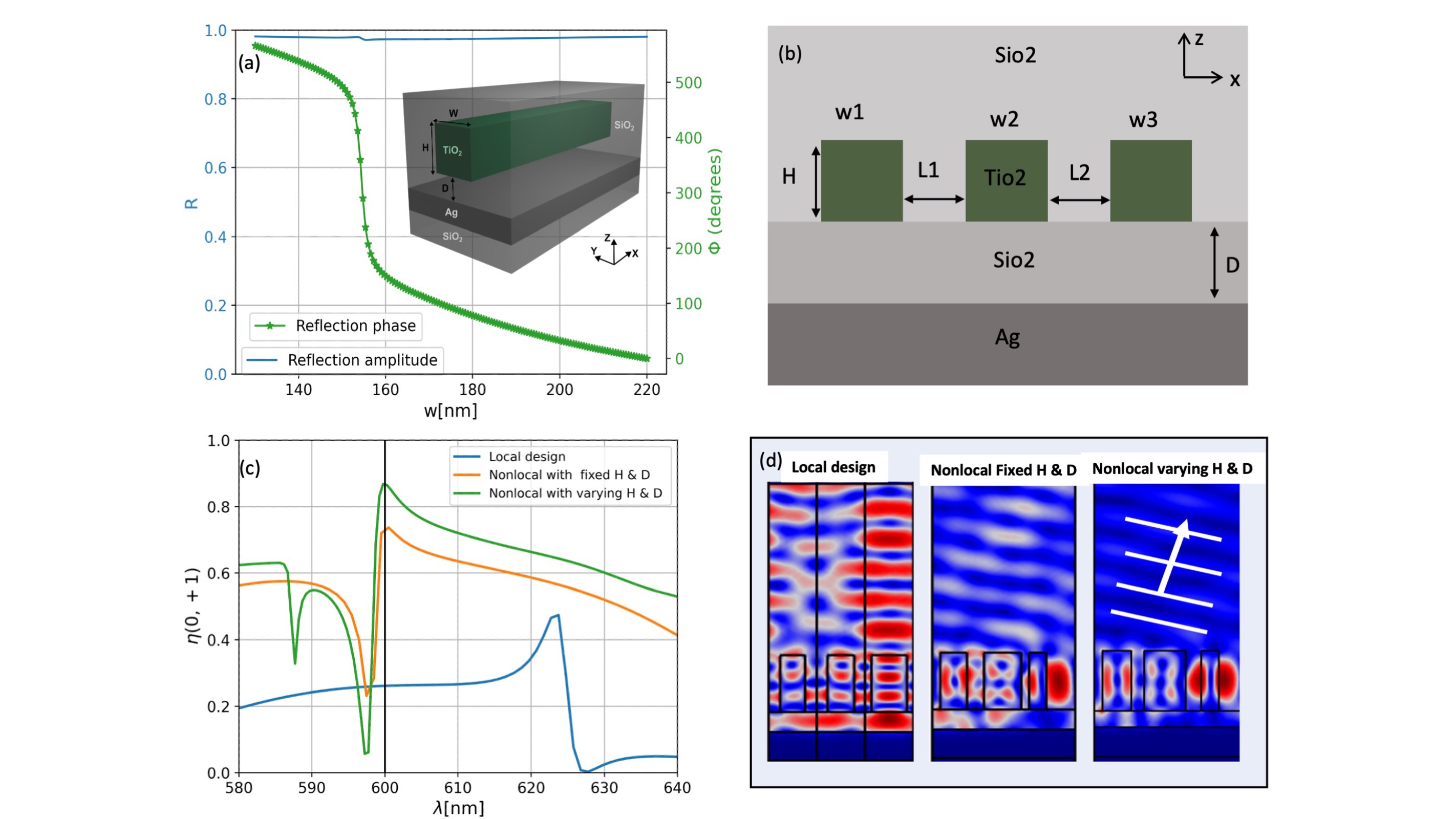}
			\caption{The performance of the nonlocal metasurface device with a nonlinear phase response and high reflection amplitude response. (a): Phase and reflection response for single unit cell simulations as a function of the width of the \ch{TiO2} nano ridges. The blue curve and left label represent the constant reflectivity of the device, while the orange curve and right label represent the nonlinear phase response of the device. The inset in (a) indicates the schematic representation of the structure under consideration. We consider a \ch{TiO2} nano ridge (invariant along $y$ direction) encapsulated in \ch{SiO2} material. The refractive index of the Ag layer is obtained from Ref. \cite{johnson1972optical}, while the refractive index of \ch{TiO2}, and \ch{SiO2} are fixed as $2.5$ and $1.45$, respectively. The height of the ridge and the period are  fixed as $H=400$ nm and $P=300$ nm, respectively. The width of the pillar can vary from $130$ nm to $220$ nm. (b) Beam steering configuration composed of 3 different ridges with varying distances in between. (c) beam steering response (the power reflected off in the first order mode) based on different optimization scenarios. The blue curve refers to the  classical local synthesizing approach  where the widths of the ridges are varying in a  gradient manner based on the look-up table given in (a), for a fixed period $P=300$ nm. The orange curve indicates the nonlocal optimization by varying the widths and distances in between, while keeping $H=400$ nm and $D=140$ nm as in (a). The green curve is for the full nonlocal optimization, where $H$ and $D$ are varying along with the widths and distances between ridges.
				(d) Field profiles for the optmized scenarios depicted in (c) at the desired wavelength $\lambda=600$ nm. The optimized parameters in each case are depicted in Tab. 3 in the the supplementary information section.}
			\label{fig2:GT}
		\end{center}
	\end{figure*}
	In this section, we aim to provide another description of a nonlocal metasurface example that cannot be designed efficiently using the lookup table synthesizing local approach. Specifically, we consider the classical 2D metagrating configuration to design a beam steering device in reflection. To achieve high performance, we rely on the asymmetric-cavity resonator, also known as the Gires-Tournois cavity, to design a reflective metasurface that supports a nonlinear phase response associated with a high reflection amplitude response.
	The single-unit cell, depicted in the inset of Fig.\ref{fig2:GT}(a), comprises a 2D rectangular \ch{TiO2} ridge that is separated from a Silver (\ch{Ag}) layer by a distance denoted as $D$. In this configuration, the distance $D$ is the crucial parameter required to achieve the ideal performance Gires-Tournois condition by manipulating the reflection zeros to obtain full phase modulation with unity reflectivity. Further details on this approach are available in \cite{elsawyGT2022}. Notably, the height of the ridge $H$ is chosen as 400 nm, while the distance from the \ch{Ag} layer is kept constant at $D=140$ nm in Fig.\ref{fig2:GT}(a). The design is configured to operate at a wavelength $\lambda=600$ nm.
	To achieve a step nonlinear phase response with more than $2\pi$ phase modulation, we vary the width of the \ch{TiO2} ridges from 130 nm to 220 nm while maintaining a constant reflectivity (see blue curve and left label in Fig.\ref{fig2:GT}(a)). Consequently, the single-unit cell design (with periodic boundary conditions on the slides) indicates that the resulting configuration displays a highly nonlinear phase response with a high reflection amplitude response, indicating that the device can be used effectively for beam steering purposes.

	We began our investigation by examining a supercell that consists of three ridges with different widths, arranged in a specific manner to steer reflected light in a \ch{SiO2} layer at a particular angle $\theta=27.37$\textdegree. The supercell configuration is depicted in Fig.\ref{fig2:GT} (b). Initially, we designed a classical local configuration, where the $H$, $D$, and period were fixed at $400$ nm, $140$ nm, and $P=300$ nm, respectively, as shown in the lookup table provided in Fig.\ref{fig2:GT}(a). The widths of the three ridges were selected to provide a full $2\pi$ phase variation in a gradient manner with a step of $120\deg$. The corresponding results are shown by the blue curve in Fig.\ref{fig2:GT}(a), where the total power reflected in the first order mode did not exceed $25\%$.
	
	However, when we considered the distances between the ridges as optimization parameters in addition to the widths, we observed a significant improvement in the performance of the metasurface. This resulted in a total reflected power in the first order mode of $75\%$ at the desired wavelength, as indicated by the orange curve in Fig.\ref{fig2:GT}(c). This huge improvement in the performance compared to the local approximation was due to the resonance behavior and the strong nonlocal interactions that were accurately accounted for by varying the distances between the ridges, as shown in the field profile given in the first two columns in Fig.\ref{fig2:GT}(d).
	Furthermore, we found that we could enhance the efficiency of the metasurface design by including both $H$ and $D$ in the nonlocal optimization. In this case, the efficiency reaches $85\%$ (see green curve and field map in Figs.\ref{fig2:GT}(c) and (d), respectively). Interestingly, we also obtain a novel set of parameters where the $D$ value became $100$ nm instead of $140$ nm, while the optimal height was $H=400$ nm. We obtain the same conclusion when we increased the number of ridges, as depicted in the supplementary information Fig.3, where the results based on the classical local lookup table provides lower performance compared to the nonlocal optimization scenario.
	
	\section{Conclusion}
	
	In conclusion, our study highlights the importance of considering nonlocal effects and near-field coupling in the design of resonant metasurfaces. The limitations of using a local single unit-cell approximation with periodic boundary conditions have been demonstrated, emphasizing the need for more accurate predictions of metasurface behavior. The nonlocal distribution of fields plays a crucial role in determining key parameters such as resonance wavelength, quality factor, and farfield response of the metasurface. Therefore, relying solely on local lookup tables for metasurface design is insufficient.
	
	To push this limit, we employed a global advanced statistical learning optimization technique, which ensure convergence to the global design using a moderate number of solver calls compared to traditional optimization algorithms. Indeed, state-of-the art optimization algorithms as evolutionary algorithm and local gradient can be utilized, but to the extent of requiring prohibitive number of simulations, getting stuck in a local design or requiring a careful starting point.
	Three different resonant metasurfaces were considered in this study. The first one operates in transmission and utilizes the properties of resonant Huygens metasurfaces. Numerical results indicate that a beam steering metasurface achieved an efficiency of 80\%, surpassing the classical design with only 23\% efficiency. Additionally, an extended depth-of-focus (EDOF) metalens was optimized, resulting in a five fold increase in focal depth and a four fold enhancement in focusing power compared to classical designs.
	
	Furthermore, the performance of wavelength-selected metagratings in reflection was investigated based on asymmetric cavity considerations. Notably, non-intuitive designs were obtained, surpassing 85\% efficiency, which far exceeds the efficiency achieved with classical gradient phase distribution.
	
	The presence of nonlocal effects arising from the interaction between neighboring unit cells significantly alters the behavior of the fields. A comprehensive understanding of this phenomenon is crucial for designing high-performance resonant metasurfaces. This knowledge empowers researchers and engineers to create metasurfaces with enhanced performance, opening avenues for novel applications in various domains.
	
	\section*{Acknowledgement}
	This work was supported by the French government, through the UCAJEDI Investments in the Future project managed by the National Research Agency (ANR) under reference number ANR-15-IDEX-01. The authors are grateful to the OPAL infrastructure and the Université Côte d’Azur’s Center for High-Performance Computing for providing resources and support. Some of the experiments presented in this paper were also carried out using the PlaFRIM experimental testbed, supported by Inria, CNRS (LABRI and IMB), Université de Bordeaux, Bordeaux INP and Conseil Régional d’Aquitaine. This project was also provided with computer and storage resources by GENCI at IDRIS thanks to the grant 2023-A0130610263 on the supercomputer Jean Zay's the CSL partition. We gratefully acknowledge the French National Association for Research and Technology (ANRT, CIFRE grant number 2022/0739) for their financial support.
	
	\section*{Author contributions}
	
	M.E. proposed the structures. E.I. and M.E. performed the DGTD simulations and the optimization of the structures. All authors analyzed the simulations and optimization results. E.I. and M.E. wrote the manuscript. All authors reviewed the manuscript.

	\section*{Competing interests}
	
	The authors declare no competing interests.
	
	\section*{Data availability}
	
	The data that support the findings of this study are available from the authors upon reasonable request.
	\clearpage
	\section{Supplementary information}
	\subsection{Geometric parameters of the beam deflectors in transmission}
	\begin{table}[ht]
		\centering
		\begin{tabular}{|c|c|c|}
			\hline
			Parameters & Local design & Optimized design \\
			\hline
			$r_1$ & $205.44$ nm & $210.20$ nm \\
			\hline
			$r_2$ & $194.24$ nm & $197.13$ nm  \\
			\hline
			$r_3$ & $188.13$ nm & $186.35$ nm\\
			\hline
			$r_4$ & $183.26$ nm & $180.82$ nm \\
			\hline
			$r_5$ & $178.38$ nm & $176.58$ nm\\
			\hline
			$r_6$ & $172.66$ nm & $173.62$ nm\\
			\hline
			$r_7$ & $161.63$ nm & $161.48$ nm\\
			\hline
			$r_8$ & $130.00$ nm & $262.87$ nm\\
			\hline
			$dx_1$ & $620.00$ nm & $603.51$ nm \\
			\hline
			$dx_2$ & $620.00$ nm & $603.12$ nm \\
			\hline
			$dx_3$ & $620.00$ nm & $601.74$ nm \\
			\hline
			$dx_4$ & $620.00$ nm & $599.24$ nm \\
			\hline
			$dx_5$ & $620.00$ nm & $605.30$ nm \\
			\hline
			$dx_6$ & $620.00$ nm & $607.26$ nm \\
			\hline
			$dx_7$ & $620.00$ nm & $604.93$ nm \\
			\hline
			
		\end{tabular}
		\caption{Geometric parameters of the beam deflector designs presented in section 2.1.}
		\label{tab:optimized_beam_deflect}
	\end{table}
	\subsection{Geometric parameters of the EDOF metalenses}
	\begin{table}[htb!]
		\centering
		\begin{tabular}{|c|c|c|c|c|c|}
			\hline
			Parameter & Parabolic & Log-asphere & Optimized fixed $h$ & Optimized varying $h$\\
			
			\hline
			$r_1$ & $177.73$ nm  & $156.39$ nm  & $144.40$ nm  & $130.00$ nm \\
			\hline                                            
			$r_2$ & $192.07$ nm  & $174.19$ nm  & $181.00$ nm  & $155.95$ nm \\
			\hline                                            
			$r_3$ & $139.56$ nm  & $182.23$ nm  & $178.93$ nm  & $171.93$ nm \\
			\hline                                            
			$r_4$ & $174.94$ nm  & $189.93$ nm  & $184.89$ nm  & $167.24$ nm \\
			\hline                                            
			$r_5$ & $184.24$ nm  & $203.07$ nm  & $188.38$ nm  & $179.67$ nm \\
			\hline                                            
			$r_6$ & $192.11$ nm  & $142.56$ nm  & $197.54$ nm  & $184.94$ nm \\
			\hline                                            
			$r_7$ & $202.64$ nm  & $168.07$ nm  & $201.83$ nm  & $189.14$ nm \\
			\hline                                            
			$r_8$ & $218.37$ nm  & $175.16$ nm  & $209.44$ nm  & $195.77$ nm \\
			\hline                                            
			$r_9$ & $232.12$ nm  & $177.87$ nm  & $138.84$ nm  & $202.92$ nm \\
			\hline                                            
			$dx_0$ & $342.67$ nm & $342.67$ nm  & $318.80$ nm  & $306.93$ nm \\
			\hline                                            
			$dx_1$ & $688.89$ nm & $688.89$ nm  & $630.13$ nm  & $600.90$ nm \\
			\hline                                            
			$dx_2$ & $688.89$ nm & $688.89$ nm  & $634.97$ nm  & $693.27$ nm \\
			\hline                                            
			$dx_3$ & $688.89$ nm & $688.89$ nm  & $655.32$ nm  & $647.38$ nm \\
			\hline                                            
			$dx_4$ & $688.89$ nm & $688.89$ nm  & $643.28$ nm  & $621.93$ nm \\
			\hline                                            
			$dx_5$ & $688.89$ nm & $688.89$ nm  & $672.79$ nm  & $510.00$ nm \\
			\hline                                            
			$dx_6$ & $688.89$ nm & $688.89$ nm  & $612.42$ nm  & $510.00$ nm \\
			\hline                                            
			$dx_7$ & $688.89$ nm & $688.89$ nm  & $670.47$ nm  & $558.42$ nm \\
			\hline                                            
			$dx_8$ & $688.89$ nm & $688.89$ nm  & $614.86$ nm  & $693.27$ nm \\
			\hline                                            
			$h$ & $170.00$ nm    & $170.00$ nm  & $170.00$ nm  & $189.49$ nm \\
			\hline
			
		\end{tabular}
		\caption{Geometric parameters of the EDOF metalens designs presented in section 2.2.}
		\label{tab:optimized_EDOF_lenses}
	\end{table}
	\clearpage
	\subsection{Geometric parameters of the beam deflectors in reflection}
	\begin{table}[htb!]
		\begin{center}
			\begin{tabular}{ |c|c|c|c|c|c|} 
				\hline 
				Parameter & Local design & Nonlocal fixed H and D & Nonlocal varying H and D \\
				\hline
				$W_1$ & $155.00$ nm & $168$ nm  & $175$ nm \\ 
				\hline
				$W_2$ & $166.24$ nm & $234$ nm & $252.5$ nm  \\ 
				\hline
				$W_3$ & $220.00$ nm  & $107$ nm & $107$ nm  \\ 
				\hline
				$H$ & $400$ nm &  $400$ nm & $405$ nm \\ 
				\hline
				$D$ & $140$ nm & $140$ nm & $110$ nm  \\
				\hline
				$L_1$ & $139$ nm &$106$ nm & $77$ nm   \\
				\hline
				$L_2$ & $107$ nm & $50$ nm & $93$ nm   \\
				\hline
			\end{tabular}
			\caption{Optimized parameters for the 3 element case depicted in Fig.3(b). The corresponding deflection efficiency together with the field profile for each case are depicted in Figs.3(c) and (d), respectively.}
			\label{tab:optimized_GT_3elemn}
		\end{center}
	\end{table}
	\begin{table}[htb!]
		\begin{center}
			\begin{tabular}{ |c|c|c|c|c|c|} 
				\hline
				Parameter & Local design & Nonlocal fixed H and D & Nonlocal varying H and D \\
				\hline
				$W_1$ & $154.90$ nm & $239$ nm  & $128$ nm \\ 
				\hline
				$W_2$ & $156.73$ nm & $105.6$ nm & $145$ nm  \\ 
				\hline
				$W_3$ & $172.98$ nm  & $110$ nm & $199$ nm  \\ 
				\hline
				$W_4$ & $213.96$ nm  & $152$ nm & $96.55$ nm  \\ 
				\hline
				$H$ & $400$ nm &  $400$ nm & $350.45$ nm \\ 
				\hline
				$D$ & $140$ nm & $140$ nm & $250$ nm  \\
				\hline
				$L_1$ & $144$ nm &$106$ nm & $84.36$ nm   \\
				\hline
				$L_2$ & $135$ nm & $120$ nm & $108$ nm   \\
				\hline
				$L_3$ & $106.5$ nm & $74.8$ nm & $98.54$ nm   \\
				\hline
			\end{tabular}
			\caption{Optimized parameters for the 4 element case depicted in Sup. Fig.3.}
			\label{tab:optimized_GT_4elemn}
		\end{center}
	\end{table}
	\clearpage
	\subsection{Supplementary Figures}
	\label{sup_info:figures}
	
	\begin{figure*}[htb!]
		\centering
		\includegraphics[width=1.0\textwidth]{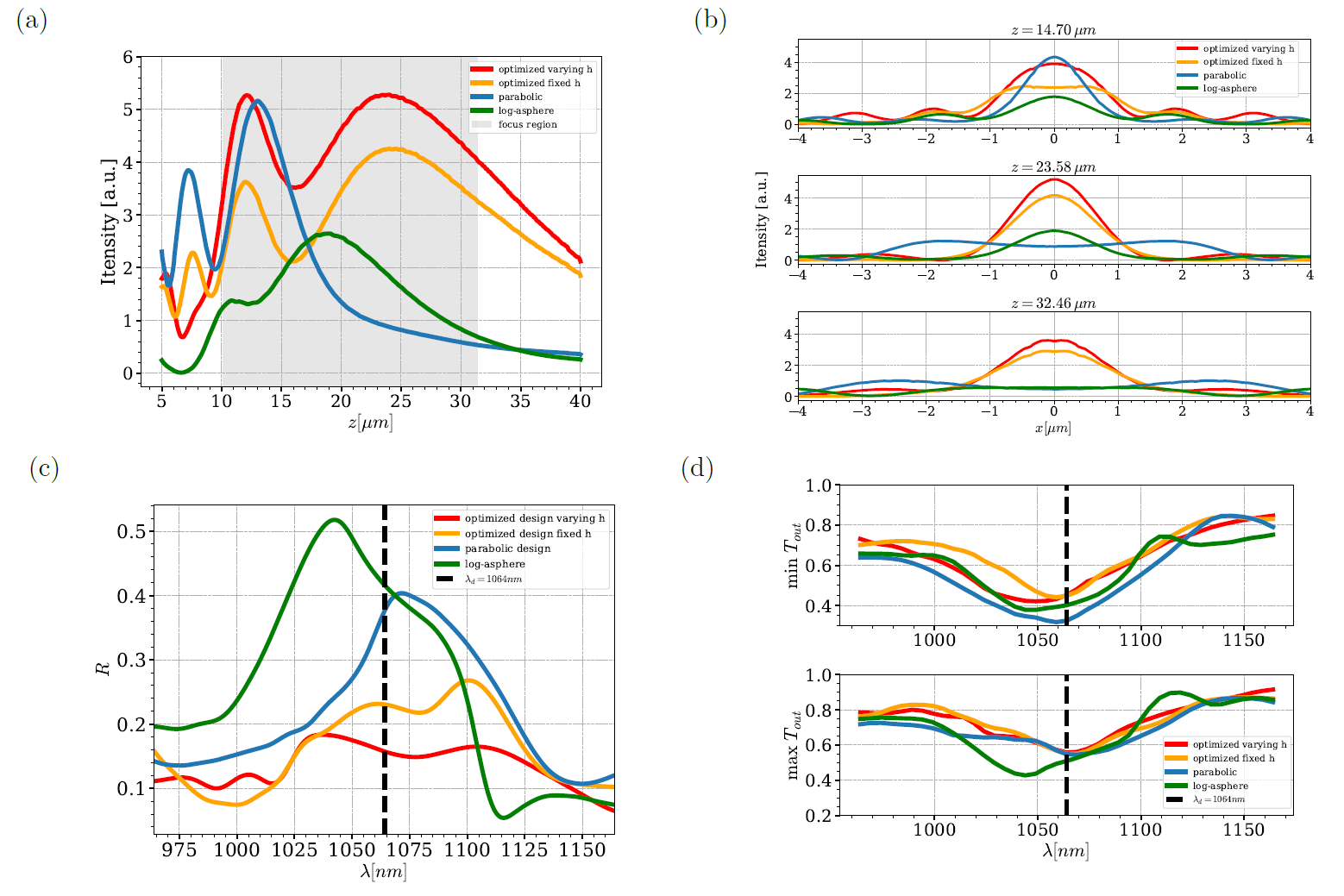}
		
		\caption{Optimization results for the EDOF metalens.(a) Intensity of the lens along the optical axis. The gray region represents the region in which we extend the depth of focus. The optimized with a varying $h$ outperforms clearly the parabolic design. The one with a fixed $h$ has a intensity profile similar to the one with varying $h$ but with consistently less intensity. (b) PSFs at $1064$nm of the different designs at $f$, $f$+2.5 DOF and $f$+5 DOF. Optimized design PSFs keep a consistent shape at the three locations while the classical designs ones have flattened considerably at $f$ + 2.5 DOF and $f$ + 5 DOF. (c) Reflection of each design. The optimized designs have considerably less back-scattering than the classical ones. Introducing $h$ as an optimization variable enables to reduce the reflection by 8\%. (d) Plot of the minimum and maximum of the part of the incident power which is scattered forward but that doesn't hit the focus surface over 20 equally spaced rectangles in the target region. It represents the large part of loose of focus efficiency. It is mainly due to the small aperture of the lens ($\sim 10 \lambda$) that causes inevitable diffraction.}
		\label{fig2:metalens2}
	\end{figure*}

	\clearpage
	\begin{figure*}[htb!]
		\centering
		\includegraphics[width=0.88\textwidth]{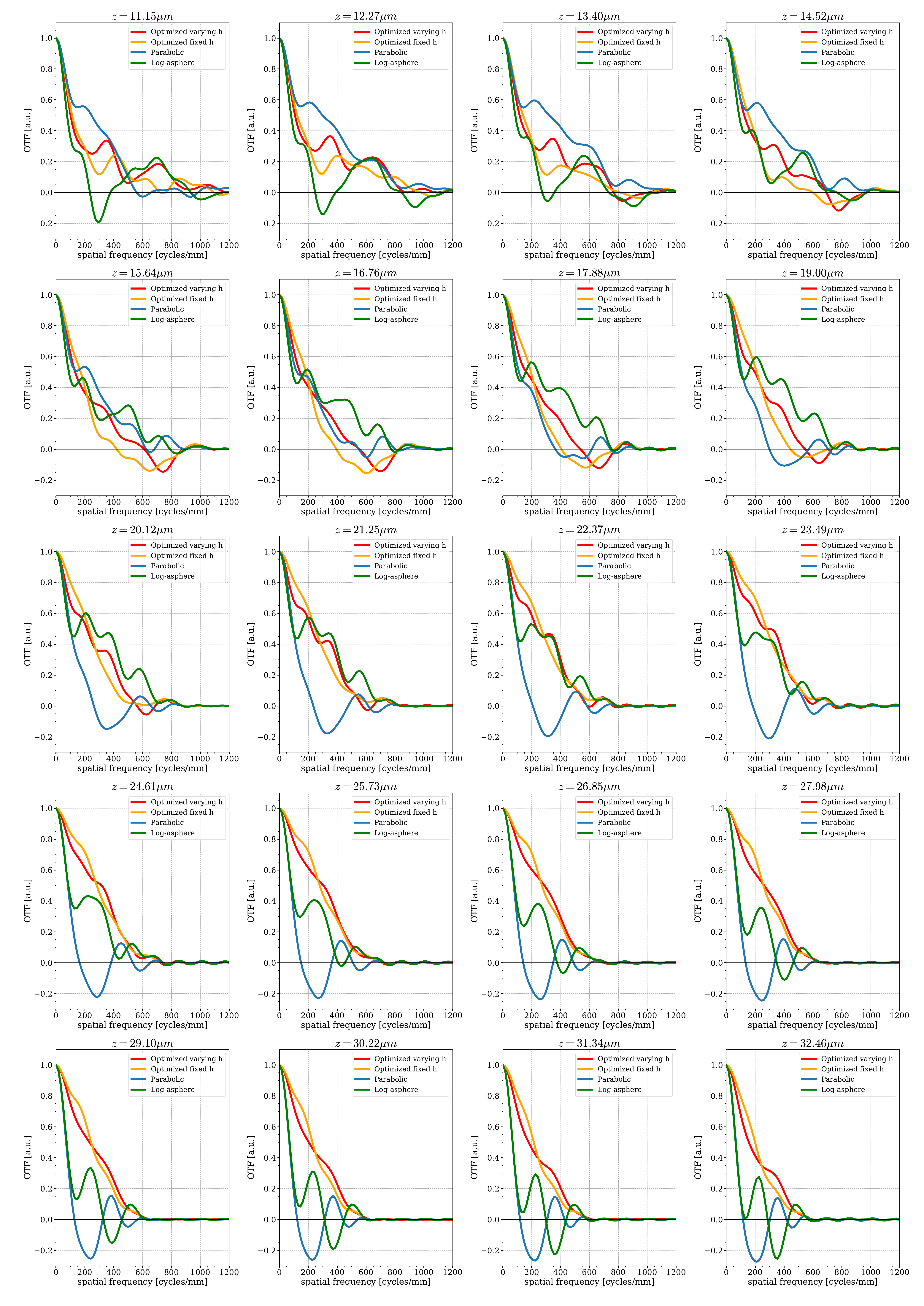}
		\caption{OTFs at 20 different focal planes equally spaced from $f$ $-$ DOF to $f$ + 5 DOF. Optimized designs exhibit a more consistent optical resolution than the classical ones.}
	\end{figure*}

	\clearpage
	
	
	\begin{figure*}[t]
		\vspace{-15cm}
		\includegraphics[width=1.0\textwidth]{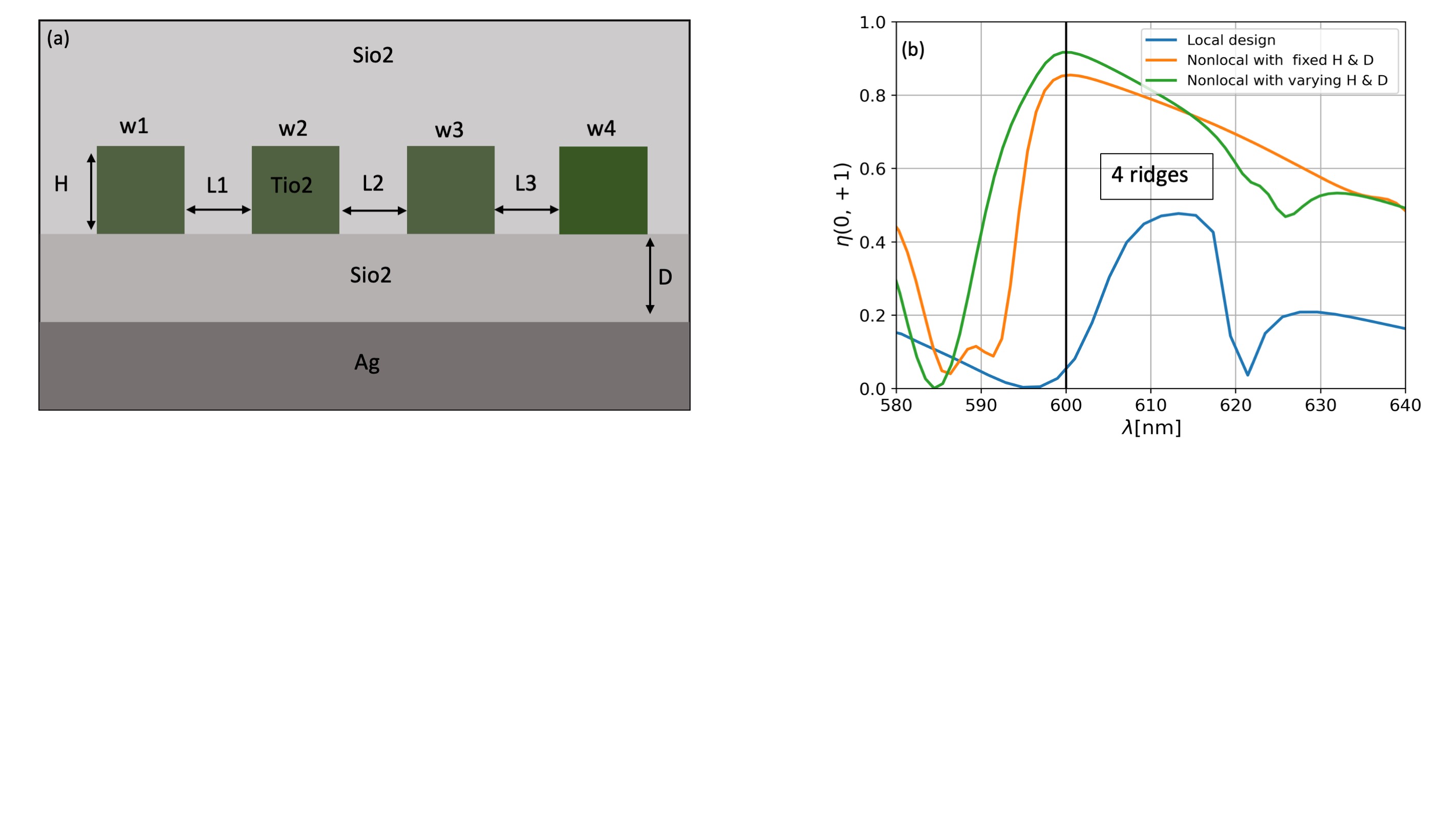}
		\caption{Optimization results for the 4 element case in reflection. (a) Geometry under consideration. (b) Deflection performance comparison between various scenarios similar to the 3 element case given in Fig.2. The optimization parameters are given in Tab. \ref{tab:optimized_GT_4elemn}.}
		\label{fig:SuppGT4ele}
	\end{figure*} 
	
	\bibliographystyle{unsrt}  
	\bibliography{nonlocal}

\end{document}